\documentclass[journal,twoside]{IEEEtran}

\usepackage{graphicx}
\usepackage{epstopdf}
\usepackage{algorithm}
\usepackage{algpseudocode}
\usepackage{enumerate}
\usepackage{amsmath}
\usepackage{multirow}
\usepackage{float}
\usepackage{array}
\usepackage{booktabs}
\usepackage{threeparttable}
\usepackage{cite}
\usepackage{color}

\ifCLASSOPTIONcompsoc
  \usepackage[caption=false,font=normalsize,labelfont=sf,textfont=sf]{subfig}
\else
  \usepackage[caption=false,font=footnotesize]{subfig}
\fi

\newcommand{\tabincell}[2]{\begin{tabular}{@{}#1@{}}#2\end{tabular}}
\begin{document}

\title{ActiveGuard: An Active DNN IP Protection Technique via Adversarial Examples}

\author{Mingfu~Xue,
        Shichang Sun,
        Can He,
        Yushu Zhang,
        Jian Wang,
        and Weiqiang~Liu

\thanks {M. Xue, S. Sun, C. He, Y. Zhang and J. Wang are with the College of Computer Science and Technology, Nanjing University of Aeronautics and Astronautics, Nanjing 211106, China (e-mail: mingfu.xue@nuaa.edu.cn; sunshichang@nuaa.edu.cn; hecan@nuaa.edu.cn; yushu@nuaa.edu.cn; wangjian@nuaa.edu.cn).}

\thanks{W. Liu is with the College of Electronic and Information Engineering, Nanjing University of Aeronautics and Astronautics, Nanjing 211106, China (e-mail: liuweiqiang@nuaa.edu.cn).}}


\maketitle

\begin{abstract}
The training of Deep Neural Networks (DNN) is costly, thus DNN can be considered as the intellectual properties (IP) of model owners.
To date, most of the existing works protect the copyright of DNN through watermarking.
However, the DNN watermarking methods can only be applied after the DNN model is stolen, which cannot resist piracy in advance.
To this end, we propose an active DNN IP protection method based on adversarial examples against DNN piracy, named \textit{ActiveGuard}.
\textit{ActiveGuard} aims to achieve authorization control and users' fingerprints management through adversarial examples, and can provide ownership verification.
Specifically, \textit{ActiveGuard} exploits the elaborate adversarial examples as users' fingerprints to distinguish authorized users from unauthorized users.
Legitimate users can enter fingerprints into DNN for identity authentication and authorized usage, while unauthorized users will obtain poor model performance due to an additional control layer.
In addition, \textit{ActiveGuard} enables the model owner to embed a watermark into the weights of DNN.
When the DNN is illegally pirated, the model owner can extract the embedded watermark and perform ownership verification.
Compared to the existing active protection works, the proposed \textit{ActiveGuard} method can realize more functions (users' fingerprints management \& copyright verification) which are required for practical commercial copyright management, while performs similar performance in active authorization control.
Experimental results show that, for authorized users, the test accuracy of LeNet-5 and Wide Residual Network (WRN) models are 99.15\% and 91.46\%, respectively, while for unauthorized users, the test accuracy of the two DNNs are only 8.92\% (LeNet-5) and 10\% (WRN), respectively.
Besides, each authorized user can pass the fingerprint authentication with a high success rate (up to 100\%).
For ownership verification, the embedded watermark can be successfully extracted, while the normal performance of the DNN model will not be affected.
Further, \textit{ActiveGuard} is demonstrated to be robust against fingerprint forgery attack, model fine-tuning attack and pruning attack.
\end{abstract}

\begin{IEEEkeywords}
Deep neural networks, active copyright protection, authorization control, users' fingerprints management, adversarial examples.
\end{IEEEkeywords}

\IEEEpeerreviewmaketitle

\section{Introduction}\label{intro}
\IEEEPARstart{D}{eep} learning techniques, especially Deep Neural Networks (DNNs), have been widely used in various fields, such as biometric authentication, speech recognition, and autonomous driving, \textit{etc}.
However, training a high-performance DNN faces the following challenges \cite{uchidanss17,rouhanick19,chenw18,abs-2011-13564}:
(i) time-consuming and requiring a large amount of training data;
(ii) expensive hardware computing resources;
(iii) requiring human experts.
Therefore, machine learning as a service (MLaaS) \cite{ribeirogc15} has emerged as a new and popular business paradigm, i.e., pre-trained DNNs are used to provide services to users.
Since the cost of stealing a pre-trained DNN model is much smaller than training a high-performance DNN model from scratch, malicious users have strong incentives to illegally copy or tamper with the DNN models \cite{chenw18,9171904}.
An adversary can easily duplicate or pirate DNN models with little prior knowledge.
Recently, the copyright protection for DNN has attracted more and more concerns.

DNN watermarking method is first proposed to protect the intellectual property (IP) of DNN.
Similar to digital watermarking in the multimedia field \cite{schyndelto94,687830}, DNN watermarking is a technique that embeds the watermarks into DNN models, and extracts the watermarks for ownership verification when copyright infringement occurs.
To date, most of the existing works protect the copyright of DNN using watermarks, which can be divided into two types, white-box watermarking methods \cite{uchidanss17,rouhanick19,nagaiuss18} and black-box watermarking methods \cite{adibcpk18,zhanggjwshm18,merrerpt20,guop18}.
However, the DNN watermarking is a passive verification method that only works after the DNN is pirated or stolen, which cannot prevent piracy in advance. Besides, it cannot provide commercial digital rights management capabilities (i.e., it can neither prevent unauthorized users from illegally accessing the DNN nor track users' identities).

To date, few works \cite{chenw18,FanNC19,0001ms20} proposed active authorization control methods to protect the copyright of DNN models, in which the authorized users can obtain a high accuracy when using DNN, while illegal users will obtain a poor accuracy.
However, the work \cite{chenw18} requires to train another anti-piracy DNN from scratch to verify whether a user is legal or not.
Besides, to use the DNN normally, an authorized user requires to pre-process each input data by a transform module, which introduces massive computational overheads.
Lastly, the DNN model in work \cite{chenw18} is trained using adversarial examples. As a result, the model only maintains high accuracy for the input adversarial examples (from authorized users), but produces low accuracy for the original clean images (from unauthorized users). However, in practice, users generally expect high accuracy on clean images.
The work \cite{FanNC19} adds a passport layer after each convolutional layer, which requires massive training time to embed the digital passport layers into the DNN.
Moreover, once an attacker obtains the original training data, he can conduct the tampering attack to modify the embedded passports, or implement reverse-engineering attacks to obtain the hidden parameters.
The hardware-assisted method in \cite{0001ms20} relies on the trusted hardware devices (as a root-of-trust) to store the key for each user, which is costly for commercial applications.
Further, these existing active authorization control methods \cite{chenw18,FanNC19,0001ms20} all lack the function of users' fingerprints management, which makes these works unsuitable for commercial digital right management applications.

In this paper, we propose an active DNN IP protection method (\textit{ActivaGuard}) via adversarial examples.
The proposed method consists of three parts, i.e., active authorization control, users' fingerprints management, and ownership verification.
To achieve authorization control, \textit{ActivaGuard} adds a control layer to DNN.
The control layer can constrain the usage of the unauthorized users on a protected DNN model, i.e., make the DNN dysfunctional to unauthorized users.
To realize users' fingerprints management, \textit{ActivaGuard} treats adversarial examples as users' fingerprints, and assigns the fingerprints to authorized users (each authorized user is assigned with an adversarial example as his fingerprint). Each authorized user can input an adversarial example into the DNN to verify his identity.
In order to realize ownership verification, \textit{ActivaGuard} embeds a numerical watermark into DNN's weights with a parameter regularizer. When the DNN is suspected to be pirated or stolen, the model owner can extract the embedded watermark for ownership verification.

The contributions of this paper are four-folds:
\begin{itemize}
\item \textbf{Active IP protection for DNN.}
We propose an active DNN IP protection method (\textit{ActivaGuard}) based on adversarial examples.
The key idea behind \textit{ActivaGuard} is that, it regards adversarial examples with specific classes and confidences as users' fingerprints, and achieves authorization control based on the uniqueness of each user's fingerprint.
The experiments on MNIST \cite{lecun1998mnist} and CIFAR-10 \cite{krizhevsky2009learning} datasets show that, for authorized users, the test accuracy of LeNet-5 \cite{726791} and Wide Residual Network (WRN) \cite{zagoruykok16} models are 99.15\% and 91.46\%, respectively.
For unauthorized users, the test accuracy of the two models are only 8.92\% (LeNet-5 \cite{726791}) and 10\% (WRN \cite{zagoruykok16}), respectively.
\item \textbf{Users' fingerprints management.}
We provide a users' fingerprints management scheme, including users' fingerprints generation, users' fingerprints allocation and users' fingerprints authentication.
The fingerprint authentication success rates among the allocated 30 authorized users range from 96\% to 100\% on MNIST \cite{lecun1998mnist} dataset, and range from 99\% to 100\% on CIFAR-10 \cite{krizhevsky2009learning} dataset.
\item \textbf{Enabling ownership verification.}
We embed a watermark into the weights of DNN by leveraging a regularizer.
The embedded watermark can be applied for ownership verification without affecting the normal usage of DNN models.
The accuracy drop of two watermarked DNNs is only -0.03\% (LeNet-5 \cite{726791}) and -0.08\% (WRN \cite{zagoruykok16}), respectively.
This indicates that the proposed \textit{ActiveGuard} method will not degrade the performance of DNN models.
Compared with existing watermarking method \cite{uchidanss17}, the proposed watermark embedding method can provide larger capacity (0-9, rather than 0/1) and is more stealthy (can be embedded discretely).
\item \textbf{Anti-attack capability.}
We evaluate the robustness of the proposed \textit{ActiveGuard} method against three common attacks.
For fingerprint forgery attack, the success rate of 10,000 forged fingerprints is between 0.01\% and 0.1\%.
Even the fine-tuning attack has been performed on the two DNNs (LeNet-5 \cite{726791} and WRN \cite{zagoruykok16}) for 50 epochs, the embedded watermark can still be extracted correctly.
For pruning attack, when 90\% parameters of the DNN are pruned, the watermark embedded by the proposed \textit{ActiveGuard} method can still be successfully extracted.
\end{itemize}

This paper is organized as follows. The related works are reviewed in Section \ref{sec2:relatedwork}.
The proposed method (\textit{ActiveGuard}), including authorization control, users' fingerprints management and copyright verification, is elaborated in Section \ref{sec3:proposedmethod}.
The effectiveness and robustness of the proposed method are evaluated in Section \ref{sec4:experiments}.
This paper is concluded in Section \ref{sec5:conclusion}.

\section{Related Work}\label{sec2:relatedwork}
In this section, we briefly review the related works about DNN IP protection, including the existing white-box watermarking methods, black-box watermarking methods and few authorization control methods.

According to the availability of DNN's parameters, DNN watermarking methods can be divided into white-box watermarking methods \cite{uchidanss17,rouhanick19}, and black-box watermarking methods \cite{adibcpk18,zhanggjwshm18,merrerpt20}.

\textbf{White-box watermarking methods.} Uchida \textit{et al.} \cite{uchidanss17} first proposed to protect the copyright of DNN by embedding a watermark.
They utilized a parameter regularizer to embed \textit{N}-bit strings into the weights of a DNN.
The embedded watermark does not affect the performance of the model, and can effectively resist model fine-tuning and pruning attacks \cite{uchidanss17}.
Rouhani \textit{et al.} \cite{rouhanick19} designed \textit{DeepSigns} to embed watermark into the probability density distribution of activation maps.
The embedded watermark can be triggered by submitting a specific input data, which can be used to verify the ownership of the DNN model remotely \cite{rouhanick19}.

\textbf{Black-box watermarking methods.} Merrer \textit{et al.} \cite{merrerpt20} embedded a watermark into the DNN through adversarial examples based stitching algorithm.
The algorithm slightly altered the decision boundary of the DNN, and the specific adversarial examples nearby the decision boundary could be used as the watermark key set to remotely verify the ownership of the model.
Adi \textit{et al.} \cite{adibcpk18} leveraged the over-parametrization of DNN models, and developed a backdoor-based watermark embedding method.
Zhang \textit{et al.} \cite{zhanggjwshm18} developed three backdoor-based watermark embedding approaches: embedding meaningful content, embedding noise, and embedding irrelevant data, as the watermark, respectively.

\textbf{Few authorization control methods.}
To date, there are three authorization control works \cite{chenw18, FanNC19, 0001ms20}.
Compared with the watermarking methods, the authorization control methods aim to provide high accuracy for authorized users, and low accuracy for unauthorized users.
Chen and Wu \cite{chenw18} proposed an adversarial example based method for authorization control.
Specifically, they trained an anti-piracy DNN module by altering the loss function of the DNN.
The anti-piracy DNN shows low accuracy for unauthorized users' inputs (raw input) and retains high accuracy for authorized users' preprocessed inputs (adversarial input).
Fan \textit{et al.} \cite{FanNC19} embedded passport layers to proactively protect the IP of DNN.
They carefully designed the passport layers to reduce the model performance when unauthorized users use the DNN, while retaining original performances for authorized users.
Chakraborty \textit{et al.} \cite{0001ms20} proposed a Hardware Protected Neural Network (HPNN) framework, which exploited the trusted hardware device as the root-of-trust for authentication to protect the IP of a DNN.
The authors leveraged a key-dependent backpropagation algorithm to train an obfuscated deep learning model.
In this way, only the authorized users with the trusted hardware device (storing the key) can run the deep learning model normally \cite{0001ms20}. The drawbacks of the three authorization control works \cite{chenw18, FanNC19, 0001ms20} have been discussed in Section \ref{intro}.

The advantages of our proposed \textit{ActiveGuard} method compared to these existing DNN IP protection methods are as follows.
\begin{enumerate}[1)]
  \item \textbf{Active authorization control and users' fingerprints management.} The existing active DNN IP protection methods \cite{chenw18, FanNC19, 0001ms20} only realize the active authorization control. To the best of the authors' knowledge, this paper is the first work which achieves active authorization control and users' fingerprints management simultaneously and these are the functions required for commercial copyright management applications.
  \item \textbf{Unique fingerprint for each authorized user.} For the authorization control method in \cite{chenw18}, each authorized user requires to preprocess each input by a conversion module to use the target DNN, which introduce high overhead. However, this paper only requires to generate one adversarial example for each user, and allocates it to each authorized user as his/her unique fingerprint. Specifically, for the first time, the proposed \textit{ActiveGuard} method utilizes the low confidence interval and generates unique adversarial example based on a specific class $t$ and an fixed confidence $c$ as a user's fingerprint.
  \item \textbf{Control layer to achieve authorization control.} The proposed method designs a control layer and adds it to the end of DNN, so as to restrict the usage of unauthorized users.
      Compared to existing active authorization control methods \cite{chenw18, FanNC19, 0001ms20} which need to train another anti-piracy DNN from scratch \cite{chenw18},
      or requires to append passport layers after each convolutional layer \cite{FanNC19},
      or requires the support of trusted hardware devices which is costly \cite{0001ms20},
      the proposed \textit{ActiveGuard} method is low-cost and low-overhead, which makes the proposed method more practical for commercial applications.
  \item \textbf{Larger capacity and more stealthy watermark embedding method.} To achieve copyright verification, the proposed \textit{ActiveGuard} embeds a numerical watermark into convolutional layer.
      Unlike the watermark in work \cite{uchidanss17} which consists of binary strings (0/1), this paper designs a watermark mapping function to map the weights of a convolutional layer to the numerical digits 0-9. In this way, for each bit of the watermark, the proposed method can embed 10 different digits (0-9) at most, which has a larger capacity than the watermarking method in work \cite{uchidanss17} (only 2 different digits at each bit). Besides, the watermark in \cite{uchidanss17} is embedded in the continuous weights of DNN. The proposed watermark can be embedded discretely among the weights of a convolutional layer, which makes the embedded watermark more stealthy.
\end{enumerate}

\section{The Proposed Method}\label{sec3:proposedmethod}
\subsection{Overall Flow}
\begin{figure*}[!htbp]
\centering
\includegraphics[width=4.8in]{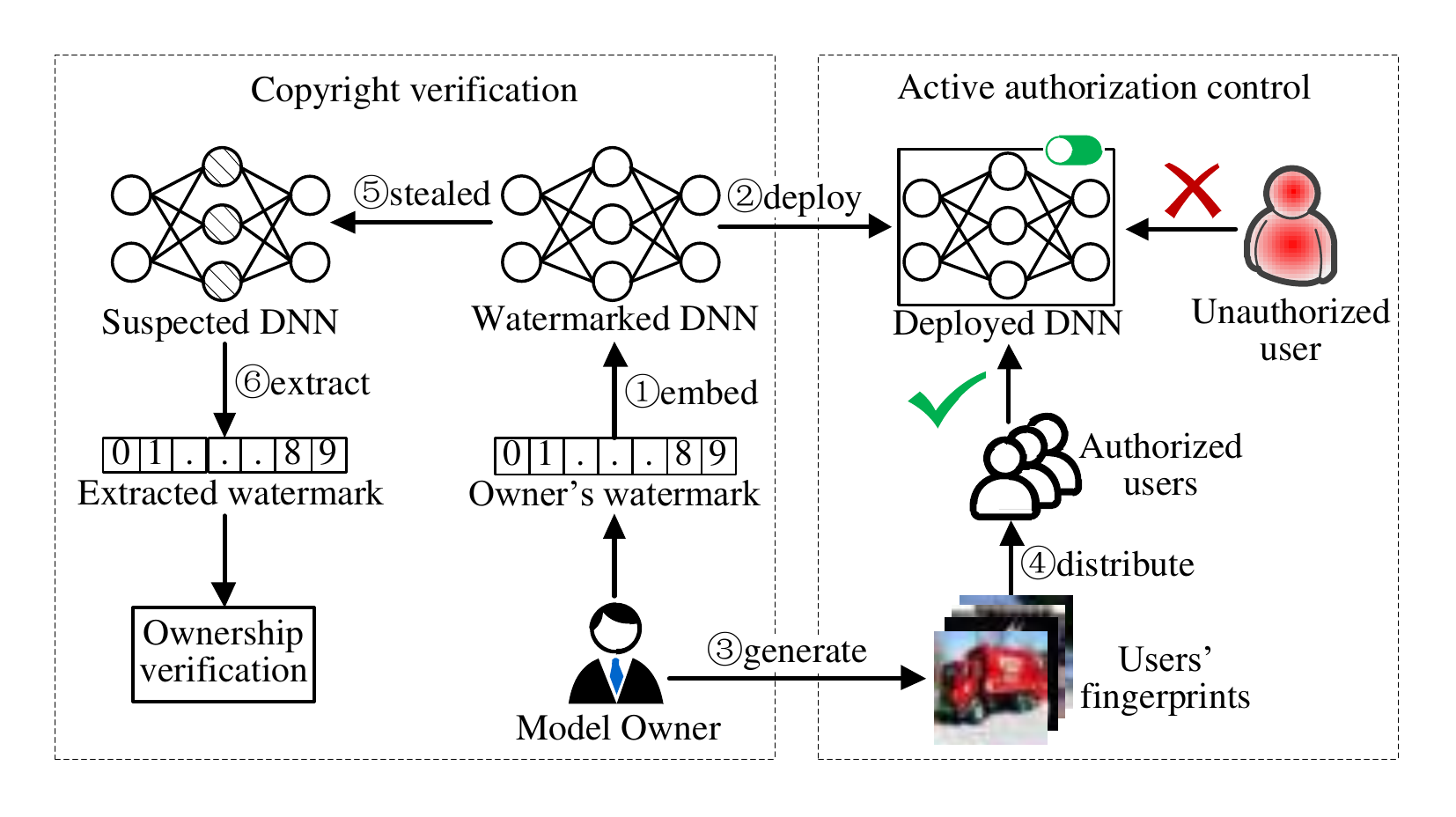}
\caption{Overview of the proposed active IP protection method (\textit{ActiveGuard}) for DNN.}
\label{fig:fig1}
\end{figure*}

In this section, we elaborate the proposed active IP protection method (\textit{ActiveGuard}) for DNN.
Fig. \ref{fig:fig1} presents the overall flow of the proposed method.
The overall flow can be divided into the following four steps:
(i) The model owner embeds the specific numerical watermark into the DNN model. The DNN with watermark is referred as watermarked DNN.
(ii) The model owner deploys the watermarked DNN as an online service, and generates the licenses (i.e., the adversarial examples) for authorized users to achieve active authorization control.
(iii) The authorized users submit fingerprints (adversarial examples) to DNN model to verify their identities, and then use the DNN normally. On the contrary, unauthorized users will obtain low performance due to the added control layer.
(iv) When the model owner suspects that the DNN has been pirated or stolen, he can extract the embedded watermark from the weights of specific convolutional layer of the suspicious DNN model.
If the watermark can be successfully extracted, the ownership of the suspicious DNN can be verified.

The proposed \textit{ActiveGuard} method has the following four functions, which will be described in Section \ref{sec3:UAC}$\sim$\ref{copy_authen}, respectively:
\begin{itemize}
  \item \textbf{Authorization control:} distinguishing the authorized users from unauthorized users.
  \item \textbf{Users' fingerprints generation:} generating fingerprints for different authorized users.
  \item \textbf{Users' fingerprints management:} distinguishing different authorized users, which includes users' fingerprints allocation and users' fingerprints authentication.
  \item \textbf{Copyright verification:} implementing the copyright verification for model owner, which includes watermark embedding, watermark extraction and verification.
\end{itemize}

\subsection{Authorization Control}\label{sec3:UAC}
\begin{figure*}[!htbp]
\centering
\includegraphics[width=6.2in]{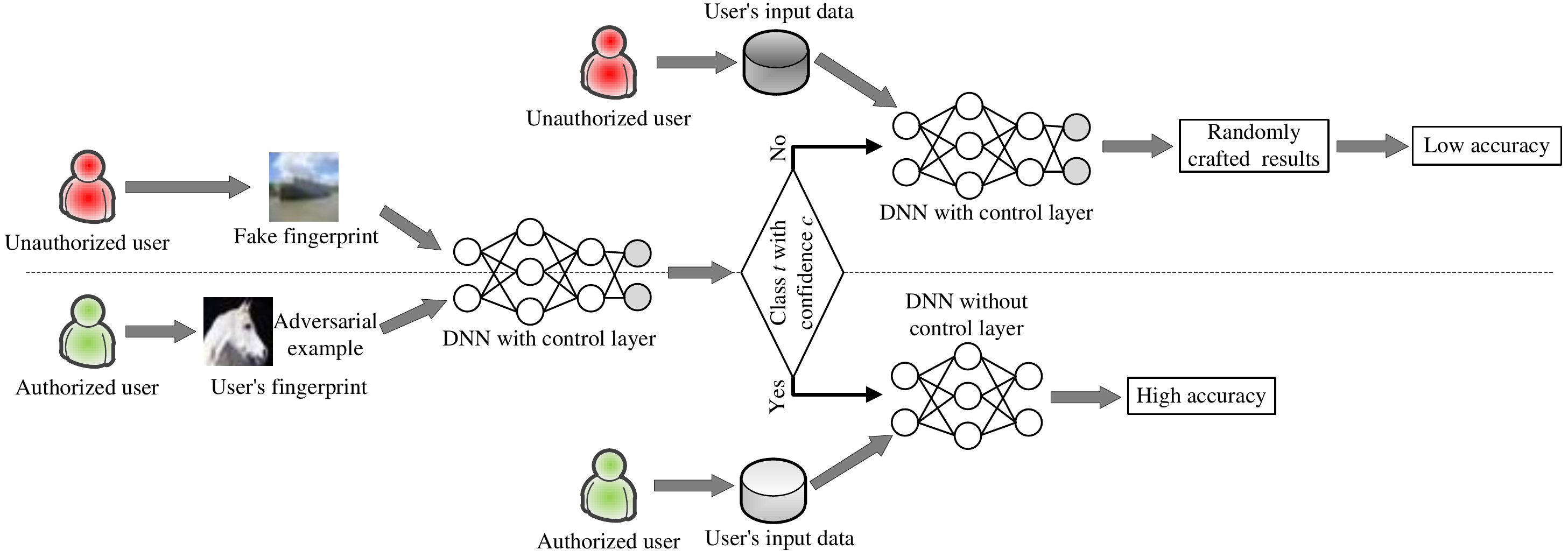}
\caption{The process of active authorization control.}
\label{fig:fig2}
\end{figure*}

First, we elaborate the process of active authorization control.
As shown in Fig. \ref{fig:fig2}, the procedure of active authorization control can be divided into three steps.
\begin{enumerate}[1)]
  \item An adversarial example is assigned to an authorized user as his fingerprint. The proposed \textit{ActiveGuard} method will generate each adversarial example based on a specific class $t$ with an fixed confidence $c$. In this way, each adversarial example is unique, thus can represent the unique identity of each authorized user.
  \item Users submit their fingerprints to DNN for identity authentication before using the DNN model.
      Specifically, we design a control layer and add it to the end of the DNN, to restrict the normal usage of unauthorized users on this protected DNN.
  \item For authorized users with legal fingerprints, the \textit{ActiveGuard} will reload the DNN model, and the added control layer will be automatically removed. In other words, the authorized users can use the DNN model normally (without the control layer).
However, for unauthorized users, the DNN with the control layer will output randomly predicted results.
\end{enumerate}

Note that, implementing the active authorization control requires to address the following three issues:
\begin{enumerate}[Q1)]
  \item How does DNN distinguish between authorized users and unauthorized users?
  \item How to generate adversarial example that represents the unique identity of each user?
  \item How does DNN distinguish between different authorized users?
\end{enumerate}

In this section, we discuss how to solve the issue Q1, the solutions to issue Q2 and Q3 will be presented in Section \ref {AEgeneration} and Section \ref {sec3:fpscheme}, respectively.

The proposed \textit{ActivaGuard} attempts to exploit the difference of confidences to distinguish authorized users from unauthorized users.
In general, for a DNN with high performance, almost all the clean inputs will be classified as ground-truth labels with high confidences.
Meantime, the well-crafted adversarial examples will be classified as their target class labels with high confidences.
In other words, almost no inputs will be classified as a class $t$ with a low confidence (below 0.50).
Inspired by the above observation, this paper utilizes the low confidence interval (ranges from 0.10 to 0.50) to achieve active authorization control.
First, we select some fixed confidences (such as 0.20, 0.30 and 0.40) from the low confidence interval $[0.10, 0.50]$, and add these selected confidences to the set $C_{fp}$, i.e., $C_{fp}=\{0.20,0.30,0.40\}$.
Second, the set $C_ {fp}$ is utilized to distinguish the authorized users from unauthorized users.
Specifically, we generate such adversarial examples that are classified by the protected DNN as the specific target classes, while their classification confidences are all in the set $C_{fp}$.
In this way, when a user submits his fingerprint (i.e., adversarial example) to the protected DNN, he would be regarded as an authorized user if his fingerprint is classified as a specific target class with a confidence in the set $C_{fp}$.
Otherwise, the user is considered to be unauthorized.

In addition, to achieve access control, we design a control layer based on the \textit{Lambda Layer} \cite{chollet2015keras}, and add the control layer to the end of the DNN.
The control layer involves several control conditions and tensor (multi-dimensional vector) operations, and the implementations of this layer are as follows:
i) Receive the predicted class and the confidence vector propagated by the output layer, and extract the highest confidence (i.e., the confidence of the predicted class) in the confidence vector.
ii) Calculate the errors between the highest confidence and each confidence in the set $C_{fp}$, and the minimal error among these calculated results is denoted as $E_{c}$.
If $E_{c}$ is less than the tolerable error, then the user is considered to be an authorized user. Otherwise, the user is considered to be an unauthorized user.
iii) Output a randomly predicted result to the unauthorized user. For authorized users, the model will be reloaded automatically and the added control layer will be removed to provide normal performance.

\subsection{Users' Fingerprints Generation}\label{AEgeneration}
Then, we introduce how to generate users' fingerprints (i.e., adversarial examples).
Formally, the protected DNN has $K$ classes and $T$ assignable confidences (i.e., the set $C_{fp}$ has $T$ elements).
In this way, a total of $K \times T$ fingerprints can be assigned to users.
The fingerprint of a user is denoted as $f$, and all the $K \times T$ fingerprints constitute the fingerprint library $FP$, i.e., $FP = \{{f_1},{f_2}, \ldots ,{f_{K \times T}}\}$.
The set of $K$ classes is represented by $L_{fp} = \{0,1,2, \ldots ,K-1\}$, and the set of $T$ confidences is denoted as $C_{fp} = \{{c_1},{c_2}, \ldots ,{c_T}\}$.

In this paper, to ensure that a generated adversarial example is unique to represent the fingerprint of each authorized user, the method to generate the adversarial examples should satisfy the following two goals:
\begin{enumerate}[1)]
\item The generated adversarial example should be classified as the target class $t$, where $t$ is a class randomly selected from the set $L_{fp}$.
\item The generated adversarial example should obtain an fixed confidence $c$ on the target class $t$ when it is input into the model $M$, where $c$ is a confidence in $C_{fp}$.
\end{enumerate}

Based on the above two goals, this paper generates the adversarial examples using the \textit{C\&W} method \cite{carlini017}. The optimization function of \textit{C\&W} method is as follows \cite{carlini017} :
\begin{equation}
\label{equ1}
  {\rm{minimize}}~ \{||\frac{1}{2}(\tanh (\delta) + 1) - x||_2^2 + \alpha  \cdot g(\frac{1}{2}(\tanh (\delta) + 1)\\\}
  \end{equation}
where $x$ is the input, $\delta$ is a variable to craft the perturbation, and $\alpha$ is a constant used to control the magnitude of the perturbation.
The first term $||\frac{1}{2}(\tanh (\delta) + 1) - x||_2^2$ is the distance function.
The second term $g(\frac{1}{2}(\tanh (\delta) + 1)$ is the objective function, where $g(\cdot)$ is originally defined as follows \cite{carlini017}:
\begin{equation}
\label{equ2}
  g_{0}(x')=\max(\max \{ Z{(x')_k}:k \ne t\} - Z{(x')_t}, 0)
\end{equation}
where $x'$ is an adversarial example, $t$ is the target class, and $Z$ is the unnormalized score from the penultimate layer of the DNN \cite{carlini017}.
In this paper, to generate users' fingerprints, a confidence control term $||Z{(x')_t} - c||$ is added to the objective function, where $c$ is an fixed confidence. Therefore, the objective function $g(\cdot)$ in this paper can be formalized as follows:
\begin{equation}
\label{equ3}
  g(x')=g_0(x') + ||Z{(x')_t} - c||
\end{equation}
The $g_0(x')$ ensures that the generated adversarial example $x'$ is classified as the class $t$.
The confidence control term $||Z{(x')_t} - c||$ guarantees that the confidence on target class $t$ is close to the predefined value $c$, where $c$ is a legal confidence in the set $C_{fp}$.

\begin{figure}[!htbp]
\centering
\subfloat[Adversarial examples generated from the MNIST dataset. The adversarial examples in the first row are classified by protected DNN as class `4' with confidence 0.40, while the adversarial examples in the second row are classified as class `5' with confidence 0.50. ]{\includegraphics[width=3.2in]{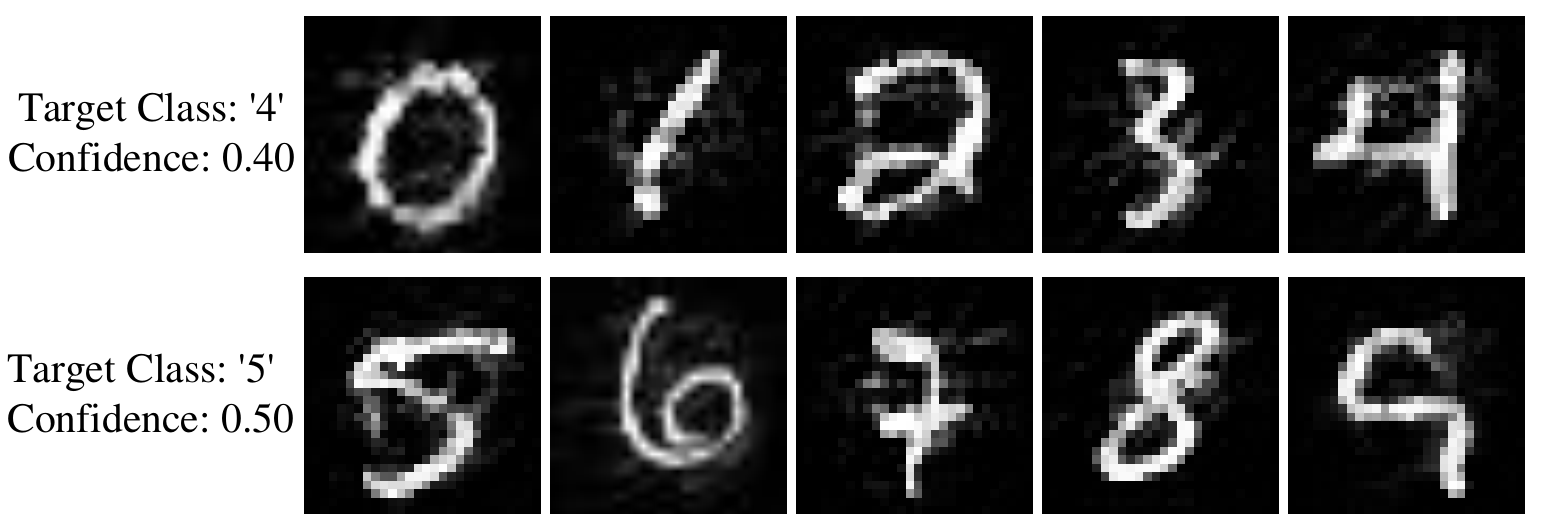}
\label{fig:fig4-a}}
\hfil
\subfloat[Adversarial examples generated from the CIFAR-10 dataset. The adversarial examples in the first row are classified as class `0' (i.e., `Airplane') with confidence 0.20, while the adversarial examples in the second row are classified as class `9' (i.e., `Truck') with confidence 0.30. ]{\includegraphics[width=3.2in]{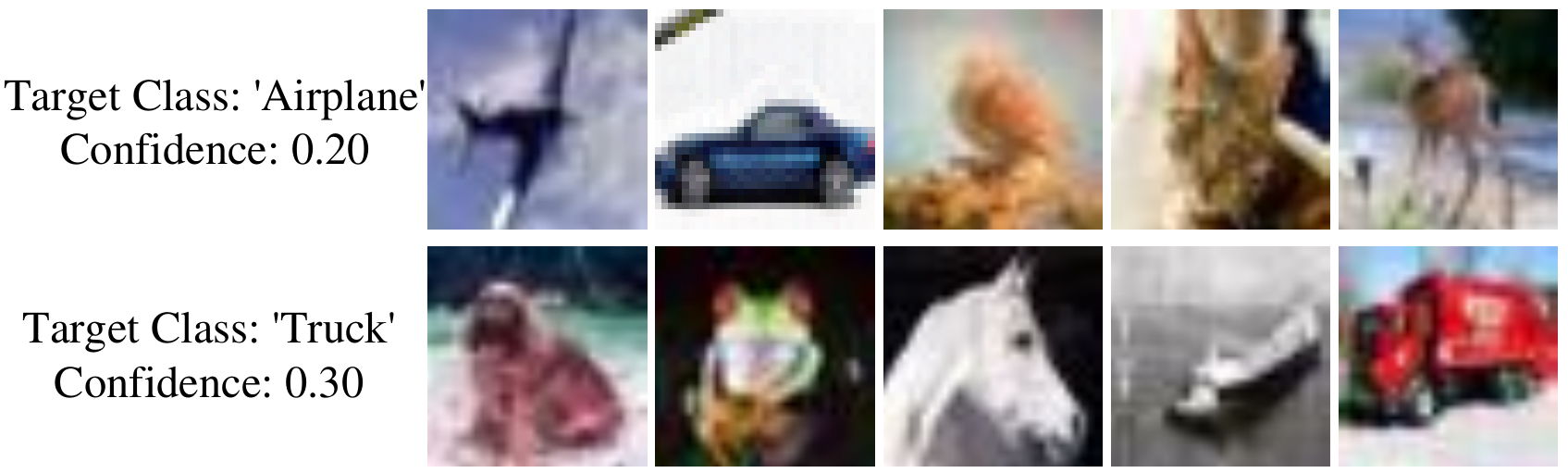}
\label{fig:fig4-b}}
\caption{Some examples of users' fingerprints that generated from MNIST and CIFAR-10 datasets.}
\label{fig:fig4}
\end{figure}

The proposed \textit{ActiveGuard} generates effective adversarial examples by minimizing the above optimization function (Equation \ref{equ1}).
Fig. \ref{fig:fig4} presents some adversarial examples that generated from the MNIST \cite{lecun1998mnist} dataset and the CIFAR-10 \cite{krizhevsky2009learning} dataset, respectively.
For MNIST dataset (as shown in Fig. \ref{fig:fig4}(a)), the target class of these generated adversarial example is `4' (the first row) and `5' (the second row), while the confidence output by the DNN model is 0.40 and 0.50, respectively.
For CIFAR-10 dataset (as shown in Fig. \ref{fig:fig4}(b)), the target class of these generated adversarial example is `0' (the first row) and `9' (the second row), while the confidence output by the DNN model is 0.20 and 0.30, respectively.

\subsection{Users' Fingerprints Management} \label{sec3:fpscheme}
Next, we discuss how to distinguish different authorized users, i.e., users' fingerprints management.
To this end, the proposed \textit{ActiveGuard} ensures that each authorized user is assigned with a unique fingerprint (an adversarial example).
The users' fingerprints management includes two parts: users' fingerprints allocation and users' fingerprints authentication, which will be discussed as follows.

\textbf{Users' fingerprints allocation.}
In \textit{ActiveGuard}, we assign a unique fingerprint to each user based on the output of DNN.
In other words, the fingerprint of a user will be uniquely determined by a class $t$ with a confidence $c$.
The legal combinations of class $t$ and confidence $c$ are denoted as \textit{Fingerprinting Output} (FO), i.e., ${\rm{FO}} = \{ (t,c)|t \in {L_{fp}},c \in {C_{fp}}\}$.
Therefore, the users' fingerprints allocation can be implemented by assigning users with different FOs, as follows:
\begin{itemize}
  \item The ${\rm{FO}}=(i,{c_j})$ is assigned to user ($i \cdot T + j$), where $i \in \{0,1, \ldots ,K-1\}$, $j \in \{1,2, \ldots ,T\}$.
\end{itemize}

\begin{figure}[!htbp]
\centering
\includegraphics[width=3.5in]{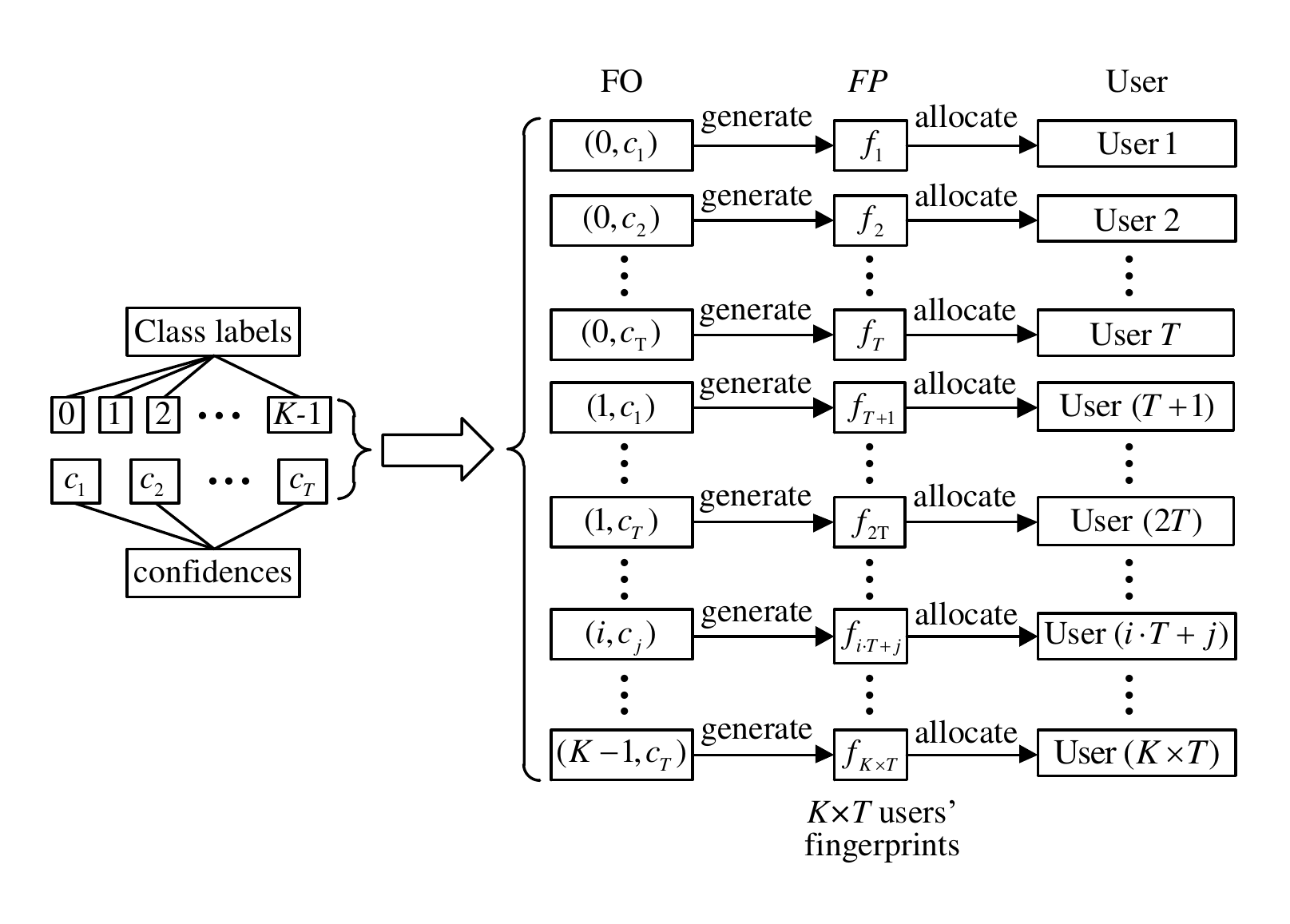}
\caption{The process of users' fingerprints allocation.}
\label{fig:fig3}
\end{figure}

The process of users' fingerprints allocation is illustrated in Fig. \ref{fig:fig3}.
First, according to all $K$ class labels and $T$ predefined confidences, a total of $K \times T$ legal FOs are obtained.
Second, the users' fingerprints are generated based on the above FOs.
Finally, the generated fingerprint $f_{i \cdot T + j}$ is allocated to an authorized user.
For the protected DNN model, the authorized user with fingerprint $f_{i \cdot T + j}$ will be classified as class $i$ with the confidence $c_j$.

\textbf{Users' fingerprints authentication.}
Based on the prediction of the model, the proposed method can:
\begin{enumerate}[i)]
  \item Determine the legitimacy of each user. Given a DNN model $M$ and an input $x$, the output confidence is denoted as a vector $P(x)$, where $P(x)$ is a $K$-dimensional vector. The output class with the highest confidence is denoted as $M(x)$.
      For a user $u$ with the fingerprint $f$, if $M(f) \in L_{fp}$ and $\max P(f) \in C_{fp}$, the user authentication would be considered to be successful, and the DNN would treat this user $u$ as an authorized user.
      For a user $\widehat u$ with the fingerprint $\widehat f$, if $M(\widehat f) \notin L_{fp}$ or $\max P(\widehat f) \notin C_{fp}$, then the authentication fails and the user $\widehat u$ would be regarded as an unauthorized user.
  \item Determine the identity of each authorized user.
      Each user's fingerprint is determined by a unique adversarial example, which is classified by the DNN model as the class $M(f)$ (i.e., target class $t$) with the confidence $\max P(f)$ (i.e., confidence $c$).
      Based on the above prediction result $(M(f), \max P(f))$, the identity of each authorized user can be determined.
\end{enumerate}

There will be a slight error between the output confidence $\max P(f)$ and $c$ during the users' fingerprints authentication, i.e., $\max P(f) \approx c$, where $c$ is a legal confidence in $C_{fp}$.
We denote the tolerable error of the confidence as $\varepsilon$.
If the confidence $\max P(f)$ output by DNN is within the error range $(c - \varepsilon ,c + \varepsilon )$, the $\max P(f)$ would be considered to be matched with $c$, i.e., the user passes the identity authentication.

\subsection{Copyright Verification}\label{copy_authen}
Finally, we introduce how to verify the ownership of the model.
Note that, the adversarial example based method is difficult to distinguish the model owner from the authorized users.
Therefore, we propose a watermarking method for copyright verification.

Inspired by the watermark embedding method in work \cite{uchidanss17}, this paper embeds a $n$-digits watermark into the weights of DNN's convolutional layer for copyright verification.
Compared to the watermarking method in \cite{uchidanss17}, our proposed watermark embedding method has two significant advantages:
i) First, the watermark in work \cite{uchidanss17} consists of binary strings (0/1), where each bit of a watermark can only be embedded with two different digits (0 or 1).
The proposed method extends the form of watermarking to numerical digits 0-9 through a linear mapping.
This allows each bit of a watermark to be embedded with 10 different digits (i.e., 0-9), which greatly improves the capacity of watermark embedding.
ii) Second, the proposed method can embed the watermark discretely, i.e., the watermark can be embedded into discontinuous weights of a DNN model, while the watermark in work \cite{uchidanss17} is embedded in consecutive positions.
As a result, our embedded watermark is more stealthy (more difficult to be noticed) and more flexible.

The process of the proposed copyright verification method includes the following two parts: watermark embedding, watermark extraction and verification.

\subsubsection{Watermark Embedding} \label{sec3:wmE}
First, to achieve copyright verification, the proposed \textit{ActiveGuard} aims to embed $n$-digits watermark ${\bf{wm}} = ({d_1},{d_2},...,{d_n})$ into the weights of the target DNN model $M$. In this paper, the watermark is embedded into a specific convolutional layer of DNN.
For a DNN, the weight matrix of a convolutional layer can be denoted as a 4-dimensional tensor ${\bf{D}} = (F,F,I,O)$, where $F$ is the size of the convolution kernel, $I$ is the number of input channels, and $O$ is the number of output channels \cite{uchidanss17,chenrfzk19}.
In this paper, the watermark is embedded into the maximum component among all $O$ components of tensor $\bf{D}$, where each component is a tensor in the form of $(F,F,I)$.
In this way, a total of $F \times F \times I$ positions are available to embed the watermark, and the weights at these $m$ ($m = F \times F \times I$) positions are denoted as a vector $\bf{w}$. We define another weight vector $\bf{v}$ to represent the weights at the $n$ ($n<m$) randomly selected positions where the watermark is embedded.

Second, different from the watermarking method in \cite{uchidanss17}, this paper aims to embed $n$-digits watermark $\bf{wm}$, i.e., the watermark consists of a series of numerical digits, which range from 0 to 9.
Generally, in order not to affect the performance of the target DNN model, the weights of DNN after embedding the watermark should be close to the original weights.
To this end, we design a watermark mapping function \textit{map($\cdot$)}, which can map the weight vector $\bf{v}$ (small values) to the watermark vector $\bf{wm}$ (large digits). In other words, the designed watermark mapping function linearly amplifies the weight values (e.g., 0.22, 0.34) to map them to watermark digits (0-9).
The \textit{map($\cdot$)} is a linear function and can be formalized as $d = ah + b$.
The $h$ is a weight value, $d$ is a digit value of the watermark, $a$ and $b$ are constants.
The steps to design the mapping function \textit{map($\cdot$)} are as follows:
\begin{enumerate}[1)]
\item Extract the weights from the original/clean DNN model and calculate the $range({\bf{w}})$ of vector $\bf{w}$.
      The $range(\cdot)$ is used to find the minimal value $w_{min}$ and maximal value $w_{max}$ in the vector $\bf{w}$, and return a range [$w_{min}$, $w_{max}$].
      For example, if ${\bf{w}}=(0.18,0.30,0.16,0.18,0.22,0.34)$, then the $range({\bf{w}})=[0.16, 0.34]$, i.e., the $w_{min}$ is 0.16 and the $w_{max}$ is 0.34.
\item Determine the constant $a$ and $b$ of function \textit{map($\cdot$)}. The \textit{map($\cdot$)} can map the above $range({\bf{w}})$ to numerical digits $[0, 9]$, i.e., \textit{map}$(range({\bf{w}})) \to [0,9]$. If $range({\bf{w}})=[0.16, 0.34]$, the constant $a$ and $b$ can be determined by solving the equation ${[0.16,0.34]^{\rm{T}}} \cdot a + b = {[0,9]^{\rm{T}}}$, and the solution of the equation is: $a=50$, $b=-8$. In this way, the watermark mapping function $map(\cdot)$ is $d=50h-8$.
\end{enumerate}

Finally, in order not to affect the performance of DNN models, the parameter regularizer (such as $L_1$-norm or $L_2$-norm) can be applied to embed the watermark \cite{uchidanss17}.
In this paper, the proposed \textit{ActiveGuard} exploits the mean square error \cite{1971Mean} to calculate the loss between the embedded watermark $\bf{wm}$ and the vector $map({\bf{v}})$.
The calculated result will be added to the original loss function of the DNN as a parameter regularizer, which aims to constrain the influence on performance that caused by the watermark embedding.
In this way, the final loss function $L$ of the target DNN model can be formalized as follows:
\begin{equation}
\label{equ4}
 L = {L_0} + \lambda \frac{1}{n}\sum\limits_{k = 1}^n {{{({d_k} - map({v_k}))}^2}}
\end{equation}
where $L_0$ is the original loss function. The ${d_k} \in {\rm{\bf{wm}}}$ is the $k$-th digit of the watermark, and ${v_k} \in {\rm{\bf{v}}}$ is the weight value in the position corresponding to the watermark digit $d_k$.
The second term $\frac{1}{n}\sum\limits_{k = 1}^n {{{({d_k} - map({v_k}))}^2}}$ is the mean square error \cite{1971Mean} between $map({\bf{v}})$ and $\bf{wm}$, and $\lambda$ is a parameter to adjust the term.

In conclusion, embedding the watermark is to modify the values of vector $\bf{w}$, so that the specific $n$-digits watermark can be embedded into the weights of a specific convolutional layer.
The overall process can be summarized as follows:
i) $n$ different positions are randomly selected from the above $m$ ($F \times F \times I$) positions in the vector $\bf{w}$ for watermark embedding. The $n$ positions are denoted as a vector $\bf{p}$ (${\bf{p}} = ({p_1},{p_2},...,{p_n})$), while the weights at the selected $n$ positions are represented by a vector ${\bf{v}} = ({v_1},{v_2},...,{v_n})$.
ii) The mean square error (the second term in Equation (\ref{equ4})) \cite{1971Mean} between the watermark vector $\bf{wm}$ and the vector $map({\bf{v}})$ is calculated.
iii) By optimizing the loss function (Equation (\ref{equ4})), the values in vector $map({\bf{v}})$ is going to approach the values in watermark vector $\bf{wm}$. In this way, the watermark $\bf{wm}$ can be successfully embedded into the weights of a specific convolutional layer of the target DNN (through a linearly enlarged mapping).

\subsubsection{Watermark Extraction and Verification} \label{sec3:wmEandV}
When the model owner suspects that his DNN is pirated, he can extract the embedded watermark from the suspected DNN and verify the ownership.

Algorithm \ref{alg1} summarizes the process of watermark extraction and verification. The Algorithm \ref{alg1} takes the suspected model $M'$, the target convolutional layer $l$, the watermark ${\bf{wm}}$, the positions $\bf{p}$ of the embedded watermark, and the watermark mapping function $map(\cdot)$ as inputs, and outputs the verified result $R$.
The steps of watermark extraction and verification are as follows.
\begin{enumerate}[Step 1.]
\item The model owner obtains the parameters of the suspected model $M'$, i.e., the target convolutional layer $l$, and the weights ${\bf{D}}_l$ of the layer $l$ (Line 1 in Algorithm \ref{alg1}).
\item The model owner utilizes the positions $\bf{p}$ of embedded watermark to extract the weights $\bf{v}$ of target convolutional layer $l$ at these corresponding positions (Line 2 in Algorithm \ref{alg1}).
\item The model owner exploits the function $map(\cdot)$ to map these extracted weights to watermark numbers, and compares the mapping result $\bf{wv_{p}}$ with his watermark $\bf{wm}$.
    If the $\bf{wv_p}$ is consistent with $\bf{wm}$, the model $M'$ is considered to be a pirate model. Otherwise, the model $M'$ is not a pirate model (Line 3-9 in Algorithm \ref{alg1}).
\end{enumerate}

Note that, in Step 3, since the values of weights are floating-point numbers, we round each digit of the $\bf{wv_p}$ to an integer.
The rounding operation can be expressed by the formula ${\bf{wv_p}} = Round({\bf{wv_p}})$.

\begin{algorithm}[htb]
\caption{Watermark extraction and verification}
\label{alg1}
\begin{algorithmic}[1]
\Require{Suspected model $M'$; target convolutional layer $l$; positions of embedded watermark $\bf{p}$; watermark mapping function $map(\cdot)$; the model owner's watermark $\bf{wm}$.}
\Ensure{Verified result $R$.}
\State ${\bf{D}}_l \leftarrow Get\_Target\_Layer\_Weights(M', l)$;
\State ${\bf{v}} \leftarrow Get\_Embedded\_Weights({\bf{D}}_l, \bf{p})$;
\State Process the weights of target convolutional layer:
\Statex \quad\ ${\bf{wv_p}} = map({\bf{v}})$;
\Statex \quad\ ${\bf{wv_p}} \leftarrow Round({\bf{wv_p}})$;
\If{$Equal({\bf{wm}},{\bf{wv_p}})==True$}
\State $R=True$;
\Else
\State $R=False$;
\EndIf\\
\Return $R$.
\end{algorithmic}
\end{algorithm}

\section{Experimental Results}\label{sec4:experiments}
In this section, we evaluate the effectiveness of the proposed \textit{ActiveGuard} method on two DNN models, i.e., the LeNet-5 \cite{726791} and the WRN \cite{zagoruykok16}.
The three functions of \textit{ActiveGuard}, authorization control, users' fingerprints management and copyright verification, are evaluated in Section \ref{Author_cont} $\sim$ \ref{cp_authen}, respectively.
Further, in Section \ref{robust}, we demonstrate that the proposed \textit{ActiveGuard} method is robust to fingerprint forgery attack and two watermark removal attacks (model fine-tuning \cite{simonyanz14a, pittarasmmp17} attack and model pruning \cite{hanptd15} attack).
Finally, we compare the proposed \textit{ActiveGuard} method with these existing active DNN IP protection methods in Section \ref{comparison}.
The experiments are implemented in Python 3.7 with Keras \cite{chollet2015keras} and Tensorflow \cite{AbadiABBCCCDDDG16} platforms.

\subsection{Experimental Setup} \label{setup}
\textbf{Datasets.} In our experiments, the LeNet-5 \cite{726791} model is trained on the MNIST \cite{lecun1998mnist} dataset, and the WRN \cite{zagoruykok16} model is trained on the CIFAR-10 \cite{krizhevsky2009learning} dataset.

- MNIST dataset \cite{lecun1998mnist}. The MNIST is a handwritten digit recognition dataset, which contains 60,000 training images and 10,000 test images.
There are a total of 10 different classes in the MNIST dataset: $\{0,1,2,...,9\}$ \cite{lecun1998mnist}.
Each image is a gray image with the size of $28 \times 28$.

- CIFAR-10 dataset \cite{krizhevsky2009learning}. The CIFAR-10 dataset consists of 50,000 training images and 10,000 test images, which contains 10 different classes: Airplane, Car, Bird, Cat, Deer, Dog, Frog, Horse, Ship, Truck \cite{krizhevsky2009learning}.
Each CIFAR-10 image is a 3-channel color image with the size of 32 $\times$ 32.

\textbf{DNN models.} The proposed \textit{ActiveGuard} aims to embed the watermark into a specific convolutional layer of the DNN.
The LeNet-5 model has 2 convolutional layers, and the WRN model has 4 convolutional layers.
Compared to the LeNet-5 model on MNIST dataset (gray, single-channel), the WRN model contains more convolutional layers and the images in CIFAR-10 dataset are more complex (color, three-channel).
Therefore, in our experiments, the LeNet-5 model is trained on MNIST dataset for 50 epochs, while the WRN model is trained on CIFAR-10 dataset for 200 epochs.
For LeNet-5 model, we use Adam \cite{kingmab14} optimizer and categorical cross-entropy loss to train the model.
For WRN model, we follow the training setting in existing works \cite{uchidanss17,zagoruykok16}, and use SGD optimizer with momentum and categorical cross-entropy loss to train the model.

\textbf{Users' fingerprints settings.} For users' fingerprints generation, we adopt the \textit{C\&W} \cite{carlini017} method to generate the adversarial examples.
For MNIST dataset, the learning rate of the optimizer is 0.005. The range of the constant $\alpha$ is (0,40), and the initial $\alpha$ is set to be 20.
For CIFAR-10 dataset, the learning rate of the optimizer is 0.001, the range of the constant $\alpha$ is (0,1), and the initial $\alpha$ is set to be 0.5.

For users' fingerprints allocation, this paper generates the fingerprints for 30 authorized users ($K=10$ and $T=3$) from MNIST and CIFAR-10 datasets, respectively.
Specifically, the 10 class labels are 0-9, and the confidence set $C_{fp}$ is \{0.20, 0.30, 0.40\}.
Note that, for the proposed \textit{ActiveGuard} method, the $K$ and $T$ can take any larger values to support more authorized users.

For users' fingerprints authentication, the proposed fingerprinting scheme can allocate fingerprints to a large number of authorized users.
A user's fingerprint is considered to be legal if the confidence of this fingerprint is between $c-\varepsilon$ and $c+\varepsilon$, where $\varepsilon$ represents the tolerable error.
In this way, the error interval of each authorized user is $2\varepsilon$.
Given the $K$ different classes, the tolerable error $\varepsilon$ and the confidence interval $(z_1, z_2)$, the total number $N_{au}$ of authorized users that our proposed \textit{ActiveGuard} method can support is calculated as follows:
\begin{equation}\label{equal5}
{N_{au}} = K \times ({z_2} - {z_1})/2\varepsilon
\end{equation}
In our experiments, the tolerable error $\varepsilon$ of the confidence is set to be 0.01.
Therefore, under the setting of $K=10$ and $\varepsilon=0.01$, if the confidence interval for authorized users is $(0.10, 0.50)$, the proposed \textit{ActiveGuard} can support up to 200 ($10 \times (0.50-0.10)/0.02$) authorized users.

\textbf{Watermarking settings.}
As discussed in Section \ref{copy_authen}, the watermark is embedded into a convolutional layer of DNN.
For LeNet-5 model and WRN model, the number of convolutional layers is 2 (conv 1 and conv 2) and 4 (conv 1, conv 2, conv 3 and conv 4), respectively, where the conv $l$ represents the $l$-th convolutional layer of the DNN.
Table \ref{tab:tab1} shows the structure and the maximum length of the embedded watermark at each convolutional layer of the two DNNs.
As discussed in Section \ref{copy_authen}, the weight structure of the weight matrix $\bf{D}$ is $(F,F,I,O)$, and the maximal watermark length is calculated by $F \times F \times I$.
For example, for the second convolutional layer (conv 2) of LeNet-5 model ($D=(5,5,6,16)$), we can embed at most 150 (5 $\times$ 5 $\times$ 6) digits into the weights.

In our experiment, the length of the watermark is set to be 13, i.e., the watermark contains 13 digits.
The watermark is embedded at the position $\bf{p}$ in the conv 2 layer, where $\bf{p}$ is randomly selected from all 150 (LeNet-5) and 576 (WRN) positions in the conv 2 layer.
For LeNet-5 model, the range of weight values in the conv 2 layer is $[0.10, 0.45]$. Therefore, the watermark mapping function is calculated by equation ${[0.10,0.45]^{\rm{T}}} \cdot a + b = {[0,9]^{\rm{T}}}$, where the solution of the above equation is $a=180/7$ and $b=-18/7$. In this way, the watermark mapping function is $d = (180/7)h-(18/7)$.
Similarly, for WRN model, the range of weight values in the conv 2 layer is $[0.20, 1.10]$, thus the watermark mapping function is calculated by equation ${[0.20,1.10]^{\rm{T}}} \cdot a + b = {[0,9]^{\rm{T}}}$. The solution of the equation is $a=10$ and $b=-2$, thus the watermark mapping function is $d=10h-2$.
The constant $\lambda$ (in Equation \eqref{equ4}) is set to be 0.01.

In addition to the above experimental settings, in Section \ref{cp_authen}, we further evaluate the effectiveness of the proposed \textit{ActiveGuard} method with respect to the following aspects: 1) the length of watermark is set to be other values; 2) the watermark is embedded into different convolutional layers.

\begin{table}[!t]
\renewcommand{\arraystretch}{1.3}
\caption{The Structure and the Maximal Length of an Embedded Watermark at Each Convolutional Layer of LeNet-5 and WRN Models}
\label{tab:tab1}
\centering
\begin{tabular}{|c|c|c|c|}
\hline
Model & \tabincell{c}{Convolutional\\ layer} & \tabincell{c}{Structure of \\weight matrix $\bf{D}$} & \tabincell{c}{Maximal length of an \\embedded watermark}\\
\hline
\multirow{2}*{LeNet-5}
  & conv 1 & (5, 5, 1, 6) & 25\\
\cline{2-4}
  & conv 2 & (5, 5, 6, 16) & 150\\
\hline
\multirow{4}*{WRN}
  & conv 1 & (3, 3, 16, 32) & 144\\
\cline{2-4}
  & conv 2 & (3, 3, 64, 64) & 576\\
\cline{2-4}
  & conv 3 & (3, 3, 128, 128) & 1152\\
\cline{2-4}
  & conv 4 & (3, 3, 256, 256) & 2304\\
\hline
\end{tabular}
\end{table}

\subsection{Authorization Control Performance} \label{Author_cont}
\begin{figure}[!htbp]
\centering
\includegraphics[width=2.7in]{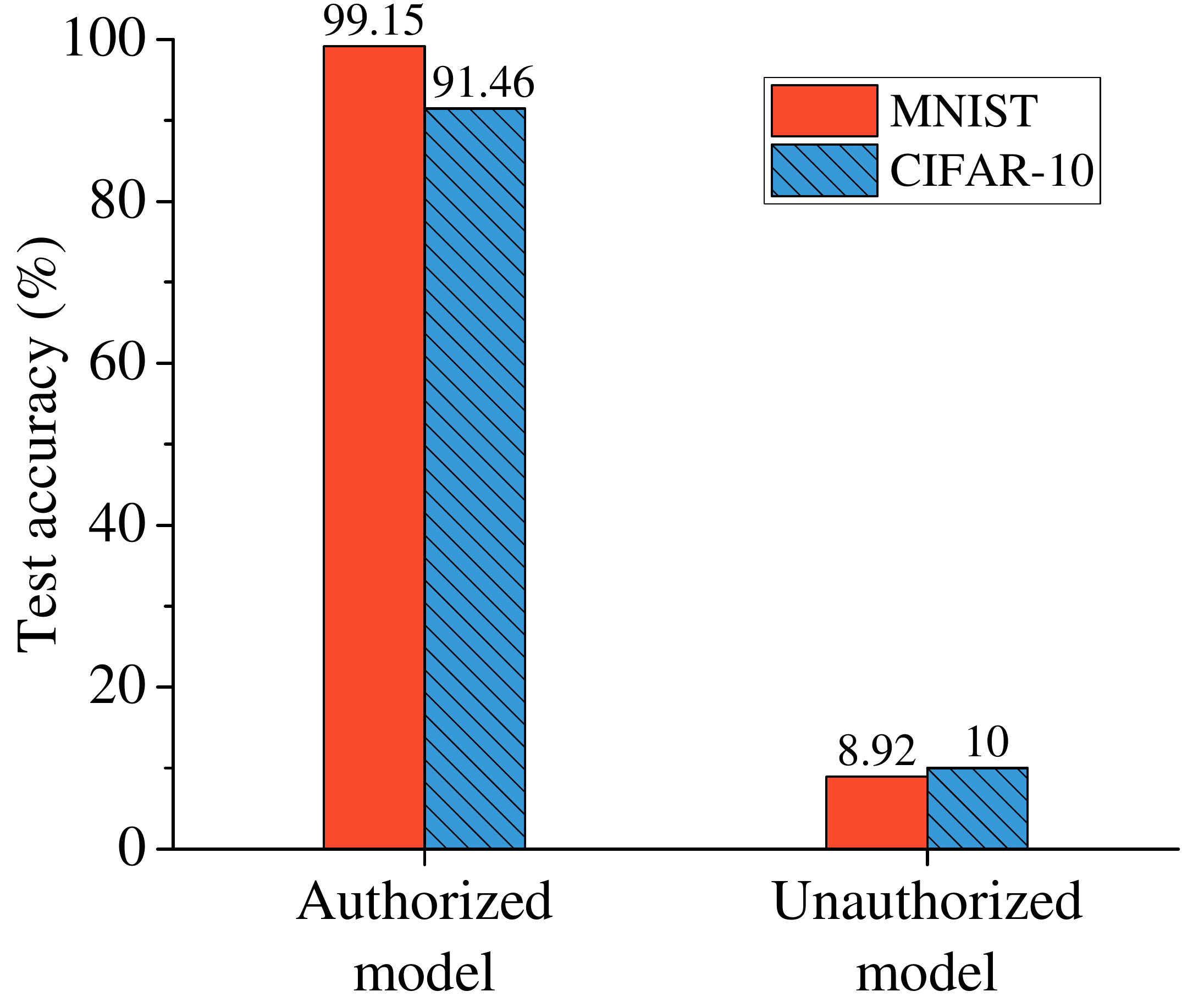}
\caption{The test accuracy of authorized model and unauthorized model on the MNIST and CIFAR-10 datasets.}
\label{fig:fig5}
\end{figure}

As discussed in Section \ref{sec3:UAC}, \textit{ActiveGuard} allows authorized users to use the DNN model normally while restricting unauthorized users.
Fig. \ref{fig:fig5} shows the test accuracy of authorized model (without the control layer) and unauthorized model (with the control layer) on two datasets.
It is shown that, the test accuracy of authorized model on two datasets is 99.15\% (MNIST) and 91.46\% (CIFAR-10), respectively.
However, the performance of unauthorized model is much lower, where the test accuracy is only 8.92\% on MNIST dataset and is only 10\% on CIFAR-10 dataset, respectively.
Therefore, the proposed \textit{ActiveGuard} method can achieve the active authorization control, and can effectively prevent the model from being used illegally.
The reason why the test accuracy for unauthorized users is close to 10\% is as follows.
As mentioned in Section \ref{sec3:UAC}, the control layer of the DNN returns randomly crafted results to unauthorized users.
This random crafting behavior is equivalent to randomly guessing a class from all $K$ classes, where the accuracy is about $\frac{1}{K} \times 100\%$ in statistics.
There are 10 ($K=10$) classes on both MNIST and CIFAR-10 datasets, thus the test accuracy for unauthorized users is around 10\%.

\subsection{Users' Fingerprints Management Performance} \label{fin_mana}
The proposed \textit{ActiveGuard} method assigns a unique adversarial example to each user as his fingerprint.
As discussed in Section \ref{setup}, there are 10 different class labels ($L_{fp} = \{0,1,2,3,4,5,6,7,8,9\}$) and 3 different confidences ($C_{fp} = \{0.20,0.30,0.40\}$).
In this way, there are a total of 30 (10 $\times$ 3) FOs.
Each FO is a combination of class label $t$ and confidence $c$, and each FO corresponds to an authorized user.
Note that, in our experiments, to evaluate the effectiveness of the proposed \textit{ActiveGuard} method, we generate 100 different fingerprints for each authorized user.
In other words, for each combination $(t,c)$, we generate 100 different adversarial examples as the fingerprints, and then submit these 100 fingerprints to DNN model to calculate the authentication success rate of an authorized user.
However, when deploys the DNNs in real world, the model owner only requires to generate one adversarial example for each authorized user, and each user performs identity authentication by submitting one fingerprint.

Table \ref{tab:tab2} reports the fingerprint authentication success rate of 30 authorized users, where the fingerprint of each user is represented by a combination $(t, c)$.
For MNIST dataset, the fingerprint authentication success rate of 30 users is 96\% at the lowest, 100\% at the highest, and 15 users achieve the success rate of 100\%.
For CIFAR-10 dataset, the minimum and maximum fingerprint authentication success rates of 30 users is 99\% and 100\% respectively, and 27 users achieve the success rates of 100\%.
It is shown that, all authorized users can successfully pass the authentication with a high success rate.
Moreover, an authorized user only requires to submit one specific adversarial example for identity authentication.

\begin{table*}[!htbp]
\centering
  \renewcommand{\arraystretch}{1.3}
  \caption{Authentication Success Rates of 30 Different Combinations of Class Label \textit{t} and Confidence \textit{c}}
  \label{tab:tab2}
    \begin{threeparttable}
    \begin{tabular}{|p{1.1cm}<{\centering}|p{1.5cm}<{\centering}|p{0.5cm}<{\centering}|p{0.9cm}<{\centering}|p{0.9cm}<{\centering}|p{0.9cm}<{\centering}|
    p{0.9cm}<{\centering}|p{0.9cm}<{\centering}|p{0.9cm}<{\centering}|p{0.9cm}<{\centering}|p{0.9cm}<{\centering}|p{0.9cm}<{\centering}|p{0.9cm}<{\centering}|}
    \hline
    \multirow{2}*{Dataset} & \multirow{2}*{Confidence \textit{c} } & \multicolumn{10}{c|}{Class label \textit{t}} \\
    \cline{3-12}          &       & 0     & 1     & 2     & 3     & 4     & 5     & 6     & 7     & 8     & 9 \\
    \hline
      \multirow{3}*{MNIST} & 0.20  & 99\%  & 99\%  & 99\%  & 99\%  & 96\%  & 100\% & 99\%  & 99\%  & 100\% & 100\% \\
    \cline{2-12}          & 0.30  & 98\%  & 100\% & 100\% & 100\% & 97\%  & 100\% & 98\%  & 98\%  & 100\% & 99\% \\
    \cline{2-12}          & 0.40  & 99\%  & 100\% & 100\% & 100\% & 97\%  & 100\% & 98\%  & 100\% & 100\% & 100\% \\
    \hline
       \multirow{3}*{CIFAR-10} & 0.20  & 100\% & 100\% & 100\% & 100\% & 100\% & 100\% & 100\% & 99\%  & 100\% & 100\% \\
    \cline{2-12}          & 0.30  & 100\% & 100\% & 100\% & 100\% & 100\% & 100\% & 100\% & 100\% & 100\% & 100\% \\
    \cline{2-12}          & 0.40  & 100\% & 99\%  & 100\% & 100\% & 100\% & 100\% & 100\% & 99\%  & 100\% & 100\% \\
    \hline
    \end{tabular}
    \begin{tablenotes}
     \item[] * Each combination of $(t,c)$ represents the identity of an authorized user.
   \end{tablenotes}
   \end{threeparttable}
\end{table*}

\subsection{Copyright Verification Performance} \label{cp_authen}
In this paper, we embed a 13-digits watermark ${\bf{wm}}_1 = [1,2,3,4,5,6,7,8,9,0,2,1,0]$ to protect the copyright of DNN model.
For watermarks with other lengths, the proposed watermark embedding method is also feasible.
Specifically, we embed the watermark with two different approaches: 1) embedding the watermark by training from scratch; 2) embedding the watermark by fine-tuning.
This paper defines a metric $V_{owner}$ to evaluate the performance of ownership verification.
If the watermark is successfully extracted from the target layer of DNN model, $V_{owner} = success$, otherwise $V_{owner} = failure$.

The results of ownership verification are shown in Table \ref{tab:tab3}.
It is shown that, the watermark can be successfully extracted to verify the ownership of the model, regardless of the two embedding approaches (training from scratch or fine-tuning).
Meanwhile, the normal performance of the watermarked DNN is similar to the performance of the DNN without watermarks.
The test accuracy of two DNNs without the watermark is 99.12\% (LeNet-5) and 91.38\% (WRN), respectively, while the test accuracy of the two watermarked DNNs is 99.15\% (LeNet-5) and 91.46\% (WRN), respectively.
This indicates that, the proposed \textit{ActiveGuard} method will not affect the normal performance of the DNN after embedding the watermark.
The reason is that, the DNN model is often over-parameterized \cite{uchidanss17,rouhanick19,chenrfzk19}, which makes the loss function of a DNN model has many local minima.
All these local minima can cause the DNN model to have the similar/good accuracy \cite{uchidanss17,rouhanick19,chenrfzk19}, thus the normal performance of the DNN model will not be affected by watermark embedding.

\begin{table*}[!htbp]
\renewcommand{\arraystretch}{1.3}
\caption{Accuracy and Ownership Verification Results on MNIST and CIFAR-10 Datasets}
\label{tab:tab3}
\centering
    \begin{tabular}{|c|c|c|c|c|c|}
    \hline
    Dataset & Model & Epoch & Test accuracy & Accuracy drop & $V_{owner}$ \\
    \hline
       \multirow{3}*{MNIST} & \multicolumn{1}{c|}{LeNet-5 without watermarks} & 50 & 99.12\% & N/A & N/A \\
    \cline{2-6}
      & \tabincell{c}{Watermarked LeNet-5 (by training from scratch)} & 50 & 99.15\% & \textbf{-0.03\%} & \textbf{success} \\
    \cline{2-6}
      & \tabincell{c}{Watermarked LeNet-5 (by fine-tuning)} & 20 & 99.12\% & \textbf{0} & \textbf{success} \\
    \hline
       \multirow{3}*{CIFAR-10} & \multicolumn{1}{c|}{WRN without watermarks} & 200 & 91.38\% & N/A & N/A \\
    \cline{2-6}
      & \tabincell{c}{Watermarked WRN (by training from scratch)} & 200 & 91.46\% & \textbf{-0.08\%} & \textbf{success} \\
    \cline{2-6}
      & \tabincell{c}{Watermarked WRN (by fine-tuning)} & 30 & 91.38\% & \textbf{0} & \textbf{success} \\
    \hline
    \end{tabular}
\end{table*}

\begin{figure}[!htbp]
\centering
\subfloat[LeNet-5]{\includegraphics[width=3.3in]{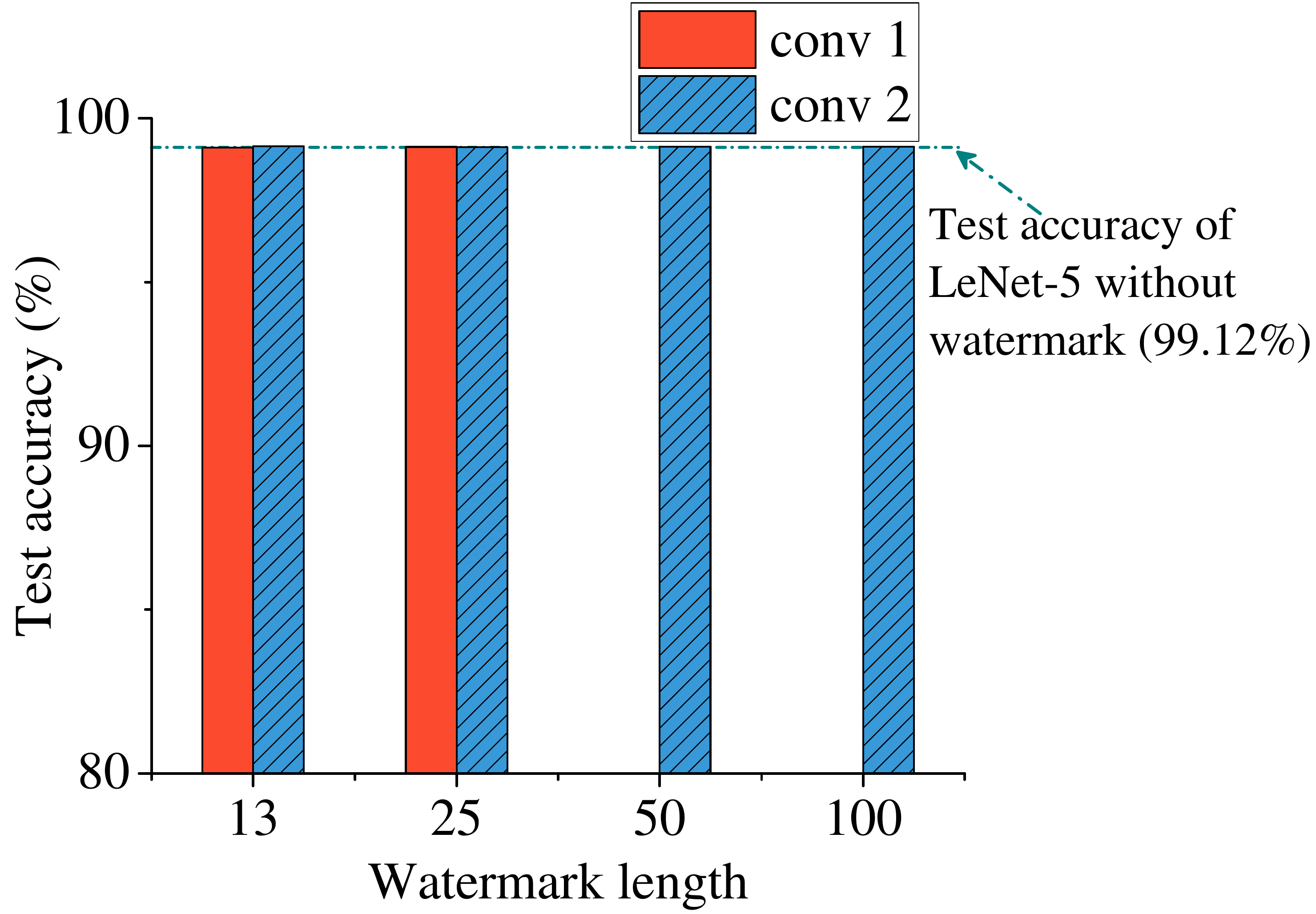}
\label{fig:fig6-a}}
\hfil
\subfloat[WRN]{\includegraphics[width=3.3in]{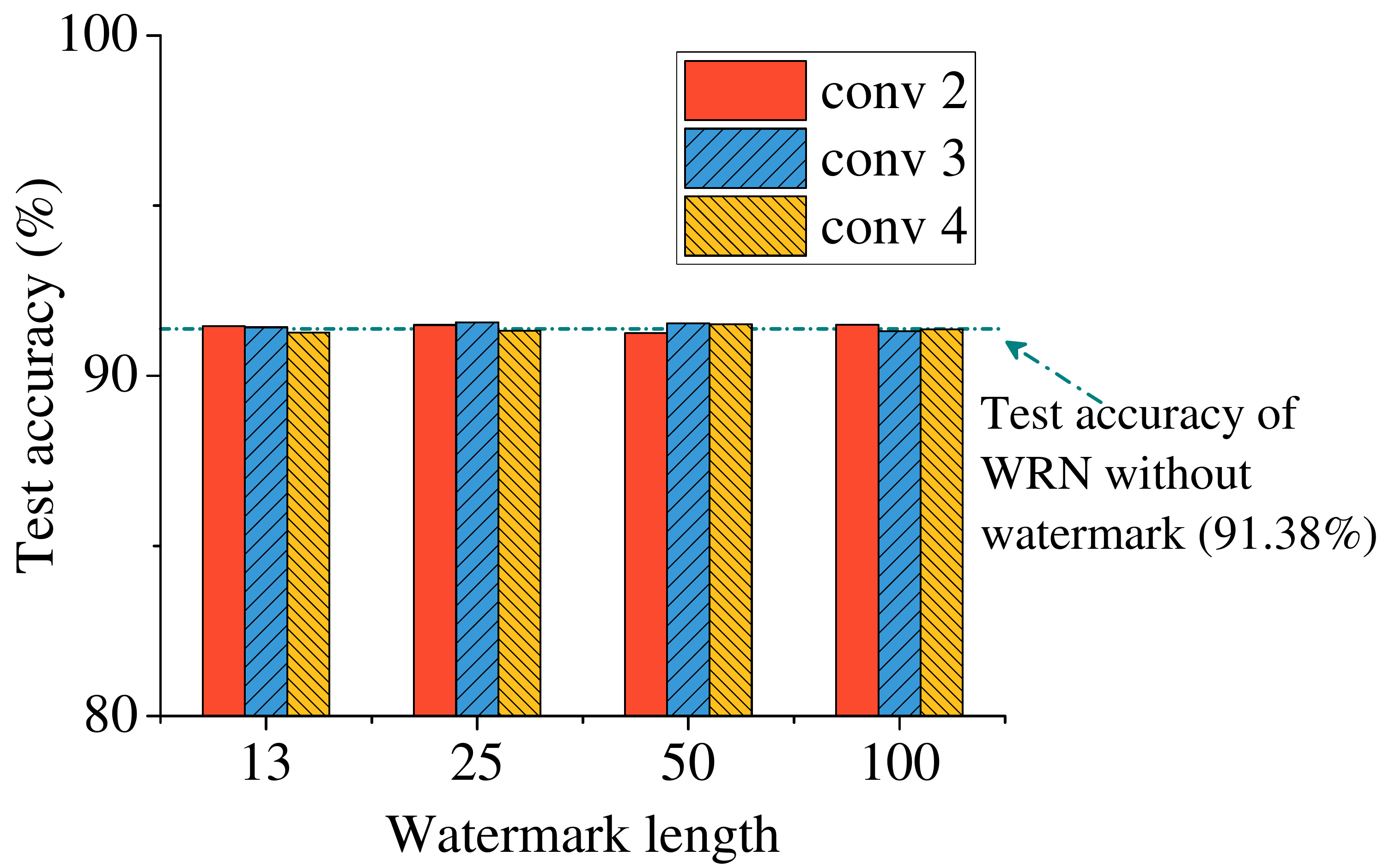}
\label{fig:fig6-b}}
\caption{The test accuracy of DNN when embedded with different lengths of watermarks into different convolutional layers.}
\label{fig:fig6}
\end{figure}

Further, we explore the impact of different watermark lengths and different target convolutional layers on the effectiveness of the proposed watermarking based copyright verification method.
Specifically, we generate the watermarks with four different lengths (13, 25, 50, 100), and embed them in the convolutional layers of LeNet-5 and WRN models, respectively.
Note that, the conv 1 layer of LeNet-5 can embed 25 digits at most, i.e., only the 13-digits and 25-digits watermarks are feasible.
The test accuracy of DNN when embedded with different lengths of watermarks into different convolutional layers is shown in Fig. \ref{fig:fig6}.
It is shown that, for LeNet-5 model, the test accuracy of the watermarked DNN ranges from 99.11\% to 99.15\%, which is consistent with the accuracy of the LeNet-5 model (99.12\%) without the watermark.
For WRN model, the test accuracy of the watermarked DNN ranges from 91.25\% to 91.57\%, which is also consistent with the test accuracy of the WRN model (91.38\%) without the watermark.
In the meantime, all the values of $V_{owner}$ are ``success'', which indicates that all the embedded watermarks can be successfully extracted for ownership verification.
Therefore, the proposed \textit{ActiveGuard} method can successfully achieve copyright verification with numerical watermarking, regardless of the lengths of watermarks and which convolutional layer the watermark is embedded.

\subsection{Robustness of the Proposed \textit{ActiveGuard} Method} \label{robust}
In this section, we evaluate the robustness of the proposed \textit{ActiveGuard} method against different attacks, including fingerprint forgery attack and two watermark removal attacks (model fine-tuning \cite{simonyanz14a, pittarasmmp17} attack and model pruning \cite{hanptd15} attack).

\textbf{Fingerprint forgery attack.}
In real world, an experienced user has strong intention to pass the DNN authentication by leveraging an forged fingerprint, which referred as the \textit{fingerprint forgery attack} in this paper.
Specifically, we consider three different fingerprint forgery attacks:
(i) using clean images as fake fingerprints ($FP_{clean}$);
(ii) fake fingerprints generated by using the \textit{FGSM} \cite{goodfellowss14} method ($FP_{FGSM}$);
(iii) fake fingerprints generated by using the \textit{C\&W} \cite{carlini017} method ($FP_{CW}$).
The \textit{FGSM} \cite{goodfellowss14} and \textit{C\&W} \cite{carlini017} methods are two popular adversarial example generation methods.
For $FP_{clean}$ attack, 10,000 clean images in the test set of MNIST (CIFAR-10) dataset are used as the fake fingerprints.
For $FP_{FGSM}$ and $FP_{CW}$ attacks, the fake fingerprints are generated by exploiting the \textit{cleverhans} \cite{papernot2018cleverhans} tool, which contains various adversarial examples generation methods.
Similarly, for $FP_{FGSM}$ and $FP_{CW}$ attacks, we select 10,000 images from the test set of MNIST (CIFAR-10) dataset to generate fake fingerprints.
The fingerprint forgery attack is considered to be successful if an forged fingerprint passes the identity authentication.
The attack success rate is the proportion of passed fake fingerprints among all generated forged fingerprints.

Table \ref{tab:tab4} shows the attack success rates of the three fingerprint forgery attacks on the MNIST and CIFAR-10 datasets. It is shown that, the attack success rates of the three forgery attacks are all lower than 0.1\%.
The reason is that, this paper exploits the adversarial example as the unique fingerprint of each user, and the fingerprint will be classified by DNN model as the target class $t$ with an fixed confidence $c$.
Therefore, for an adversary who doesn't know the protection mechanism and parameter settings at all, he cannot construct such an user's fingerprint that satisfies the above condition.
Therefore, the success rate of fingerprint forgery attack is extremely low.
In conclusion, the proposed \textit{ActiveGuard} method can effectively resist different fingerprint forgery attacks.

\begin{table}[htbp]
\renewcommand{\arraystretch}{1.3}
\caption{Attack Success Rates of Three Fingerprint Forgery Attacks on MNIST and CIFAR-10 Datasets}
\label{tab:tab4}
  \centering
    \begin{tabular}{|c|c|c|}
    \hline
    Dataset & Fingerprint forgery attack & Attack success rate \\
    \hline
    \multirow{3}*{MNIST} & {$FP_{clean}$ attack} & \textbf{0.01\%} \\
    \cline{2-3}          & {$FP_{FGSM}$ attack} & \textbf{0.10\%} \\
    \cline{2-3}          & {$FP_{CW}$ attack} & \textbf{0.01\%} \\
    \hline
    \multirow{3}*{CIFAR-10} & {$FP_{clean}$ attack} & \textbf{0.01\%} \\
    \cline{2-3}          & {$FP_{FGSM}$ attack} & \textbf{0.02\%} \\
    \cline{2-3}          & {$FP_{CW}$ attack} & \textbf{0.05\%} \\
    \hline
    \end{tabular}
\end{table}

\textbf{Watermark removal attacks.} Further, we evaluate the robustness of the proposed \textit{ActiveGuard} method against two watermark removal attacks: model fine-tuning \cite{simonyanz14a, pittarasmmp17} attack and model pruning \cite{hanptd15} attack.

\textbf{-Model fine-tuning \cite{simonyanz14a, pittarasmmp17} attack.} Fine-tuning is a conventional operation in model transfer learning.
The adversaries can exploit the fine-tuning attack to generate a new model \cite{simonyanz14a, pittarasmmp17}.
In our experiments, we select 7,000 images from the test set of MNIST dataset as the training data to fine-tune the watermarked LeNet-5, and select 7,000 images from the test set of CIFAR-10 dataset as the training data to fine-tune the watermarked WRN.
The fine-tuning attack is performed on each watermarked DNN (LeNet-5 \cite{726791} and WRN \cite{zagoruykok16}) for 30 and 50 epochs, respectively.

Table \ref{tab:tab5} shows the test accuracy and ownership verification of watermarked DNN under the fine-tuning attack.
It is shown that, for LeNet-5 and the WRN models, the watermark cannot be removed by fine-tuning attack.
Besides, the test accuracy of the watermarked DNN before and after fine-tuning attack is consistent.
For example, the test accuracy of watermarked WRN without attack is 91.46\%.
After 50 epochs of fine-tuning attack, the test accuracy of watermarked WRN remains 91.53\%.
Therefore, the proposed \textit{ActiveGuard} method is robust against the fine-tuning attack on LeNet-5 and WRN models.

\begin{table}[!htbp]
  \centering
  \renewcommand{\arraystretch}{1.3}
  \caption{Test Accuracy and Ownership Verification of Watermarked DNN under Fine-Tuning Attack}
    \begin{tabular}{|c|c|c|c|c|}
    \hline
    Model & Attack & Epoch & Test accuracy & $V_{owner}$ \\
    \hline
    \multirow{3}*{LeNet-5} & None & 50 & 99.15\% & \textbf{success} \\
    \cline{2-5}
      & \tabincell{c}{Fine-tuning attack} & 30 & 99.53\% & \textbf{success} \\
    \cline{2-5}
      & \tabincell{c}{Fine-tuning attack} & 50 & 99.53\% & \textbf{success} \\
    \hline
    \multirow{3}*{WRN} & None & 200 & 91.46\% & \textbf{success} \\
    \cline{2-5}
      & \tabincell{c}{Fine-tuning attack} & 30 & 91.47\% & \textbf{success} \\
    \cline{2-5}
      & \tabincell{c}{Fine-tuning attack} & 50 & 91.53\% & \textbf{success} \\
    \hline
    \end{tabular}
  \label{tab:tab5}
\end{table}

\textbf{-Model pruning \cite{hanptd15} attack.}
Pruning is a widely used operation for model compression, which aims to prune the DNN by removing redundant parameters from the neural network \cite{hanptd15}.
Therefore, an adversary can perform pruning attack to remove the embedded watermark in a DNN model.
In our experiments, we adopt the pruning method in work \cite{hanptd15} to prune the watermarked DNNs (LeNet-5 and WRN), i.e., the weights with small values are set to be 0 \cite{hanptd15}.
As discussed in Section \ref{cp_authen}, the watermark embedded into the LeNet-5 and WRN models is ${\bf{wm}}_1 = [1,2,3,4,5,6,7,8,9,0,2,1,0]$.

The robustness of the watermarked DNN against the pruning attack is shown in Table \ref{tab:tab6}.
First, as the pruning rate increases, the test accuracy of LeNet-5 and WRN models both decreases sightly.
Even 60\% of the weights in watermarked DNN are pruned, the test accuracy of two models still remains 98.54\% (LeNet-5) and 89.87\% (WRN), respectively.
Besides, the embedded watermark performs well in terms of ownership verification.
Specifically, even 90\% weights are pruned, the watermark embedded into the WRN model can still be successfully extracted.
However, the ownership verification fails when 50\% weights in LeNet-5 model are pruned.
The reason is that, the LeNet-5 model is smaller with fewer parameters, therefore, when pruning, the smaller values in the watermark (e.g., the weights corresponding/map to watermark digits 0, 1, 2) are more likely to be pruned. As a result, the watermark is affected when the pruning rate increases to 50\%.
Since the weights with small values will be pruned in the model pruning attack, we also evaluate the robustness of the watermarked LeNet-5 mdoel against the pruning attack with a watermark with larger values. Specifically, the watermark ${\bf{wm}}_2 =[3,8,7,6,8,7,6,9,9,4,8,6,5]$ is embedded into the LeNet-5 model for evaluation.
As shown in Table \ref{tab:tab6}, for LeNet-5 model embedded with $\bf{wm}_2$, even 90\% weights are pruned, the watermark $\bf{wm}_2$ embedded into the model can still be successfully extracted. The reason is that, in model pruning attack, the weights with small values are set to be 0, therefore, a watermark with large values will not be pruned. In fact, when 90\% weights are pruned, only the weights corresponding/map to watermark digits 0, 1, 2 are pruned, while the embedded watermark $\bf{wm}_2$ is not affected. Therefore, even for a small DNN model, embedding a watermark that does not contain small values (e.g., 0, 1, 2) can effectively resist pruning attack.
Overall, the proposed \textit{ActiveGuard} method is effective and robust against the pruning attack.

\begin{table*}[!htbp]
  \centering
  \renewcommand{\arraystretch}{1.3}
  \begin{threeparttable}[b]
  \caption{Test Accuracy and Ownership Verification of Watermarked DNN under Pruning Attack}
    \begin{tabular}{|c|c|c|c|c|c|c|}
    \hline
    \multirow{2}[0]{*}{Pruning rate} & \multicolumn{2}{c|}{LeNet-5 (embedded with ${\bf{wm}}_1$)} &  \multicolumn{2}{c|}{LeNet-5 (embedded with  ${\bf{wm}}_2$)} & \multicolumn{2}{c|}{WRN (embedded with ${\bf{wm}}_1$)} \\
    \cline{2-7}
     & Test accuracy & $V_{owner}$ & Test accuracy & $V_{owner}$  & Test accuracy & $V_{owner}$ \\
    \hline
    0\%   & 99.15\% & \textbf{success} & 99.16\% & \textbf{success} & 91.46\% & \textbf{success} \\
    \hline
    10\%  & 99.15\% & \textbf{success} & 99.15\% & \textbf{success} & 91.40\% & \textbf{success} \\
    \hline
    20\%  & 99.15\% & \textbf{success} & 99.09\% & \textbf{success} & 91.30\% & \textbf{success} \\
    \hline
    30\%  & 99.07\% & \textbf{success} & 99.08\% & \textbf{success} & 90.99\% & \textbf{success} \\
    \hline
    40\%  & 98.97\% & \textbf{success} & 99.06\% & \textbf{success} & 91.06\% & \textbf{success} \\
    \hline
    50\%  & 98.89\% & failure & 99.00\% & \textbf{success} & 90.28\% & \textbf{success} \\
    \hline
    60\%  & 98.54\% & failure & 98.93\% & \textbf{success} & 89.87\% & \textbf{success} \\
    \hline
    70\%  & 95.80\% & failure & 98.75\% & \textbf{success} & 88.03\% & \textbf{success} \\
    \hline
    80\%  & 91.02\% & failure & 97.46\% & \textbf{success} & 80.40\% & \textbf{success} \\
    \hline
    90\%  & 68.38\% & failure & 85.36\% & \textbf{success} & 48.35\% & \textbf{success} \\
    \hline
    \end{tabular}
    \begin{tablenotes}
     \item[] * ${\bf{wm}}_1 = [1,2,3,4,5,6,7,8,9,0,2,1,0]$, ${\bf{wm}}_2 = [3, 8, 7, 6, 8, 7, 6, 9, 9, 4, 8, 6, 5]$
   \end{tablenotes}
  \label{tab:tab6}
  \end{threeparttable}
\end{table*}

\subsection{Comparison with Existing Active DNN IP Protection Methods} \label{comparison}
Finally, we compare the proposed \textit{ActiveGuard} method with the existing active DNN IP protection works. So far, there are only three authorization control works \cite{chenw18, FanNC19, 0001ms20}. The comparisons are conducted with respect to the following three aspects: active authorization control, users' fingerprints management and ownership verification. The comparison results are presented in Table \ref{tab:compare}.
As a baseline, the test accuracy of these DNN models without any IP protection techniques are high (about 90\% or even higher) in the existing works (\cite{chenw18, FanNC19, 0001ms20}) and this paper.

\begin{table*}[htbp]
  \centering
  \renewcommand{\arraystretch}{1.3}
  \caption{Comparison Results of Proposed \textit{ActiveGuard} with Existing Active DNN IP Protection Methods}
    \begin{tabular}{|c|c|c|c|c|c|c|}
    \hline
    \multicolumn{1}{|c|}{\multirow{2}*{\tabincell{c}{Active DNN IP \\protection works}}} & \multicolumn{1}{c|}{\multirow{2}*{Datasets }} & \multirow{2}*{\tabincell{c}{Test accuracy of DNN \\without IP protection techniques}} & \multicolumn{2}{c|}{Active authorization control (accuracy)} & \multicolumn{1}{c|}{\multirow{2}*{\tabincell{c}{Users' fingerprints \\management}}} & \multicolumn{1}{c|}{\multirow{2}*{\tabincell{c}{Ownership \\verification}}} \\
\cline{4-5}          & \multicolumn{1}{c|}{} &       & Authorized users & Unauthorized users &       &  \\
        \hline
    \multirow{2}*{\cite{chenw18}} & \multicolumn{1}{c|}{MNIST } & 99.12\% & 99.23\% & 0.23\% & \multirow{2}*{NO} & \multirow{2}*{NO} \\
\cline{2-5}          & \multicolumn{1}{c|}{CIFAR-10} & 90.74\% & 90.61\% & 0.78\% &       &  \\
    \hline
    \multirow{2}*{\cite{FanNC19}} & \multicolumn{1}{c|}{\multirow{2}*{CIFAR-10}} & \multirow{2}*{91.12\%} & \multirow{2}*{90.89\%} & \multirow{2}*{around 10\%} & \multirow{2}*{NO} & \multirow{2}*{YES} \\
          & \multicolumn{1}{c|}{} &       &       &       &       &  \\
    \hline
    \multirow{2}*{\cite{0001ms20}} & \multicolumn{1}{c|}{Fashion MNIST} & 89.93\% & around 89.93\% & 10.05\% & \multirow{2}*{NO} & \multirow{2}*{NO} \\
\cline{2-5}          & \multicolumn{1}{c|}{CIFAR-10 } & 89.54\% & around 89.54\% & 9.37\% &       &  \\
    \hline
    \multirow{2}*{\textit{ActiveGuard}} & MNIST & 99.12\% & \textbf{99.15\%} & \textbf{8.92\%} & \multirow{2}*{\textbf{YES }} & \multirow{2}*{\textbf{YES}} \\
\cline{2-5}          & CIAFR-10 & 91.38\% & \textbf{91.46\%} & \textbf{10\%} &       &  \\
    \hline
    \end{tabular}
  \label{tab:compare}
\end{table*}

\textbf{Active authorization control.} The proposed \textit{ActiveGuard} method and the existing active DNN IP protection methods can all achieve the function of active authorization control. Specifically, in this paper, for authorized users, the test accuracy of two protected DNNs (LeNet-5 \cite{726791} and WRN \cite{zagoruykok16}) is high up to 99.15\%, which is higher than the existing works \cite{FanNC19, 0001ms20}, and is similar to the work \cite{chenw18}.
In the meantime, for unauthorized users, the test accuracy of two protected DNNs (LeNet-5 \cite{726791} and WRN \cite{zagoruykok16}) is as low as 8.92\%, which is similar to the works \cite{FanNC19, 0001ms20}.
However, our proposed \textit{ActiveGuard} does not require additional training of the model and does not change the parameters of the DNN for authorization control, while the works \cite{FanNC19, 0001ms20} need to retrain the DNN model, and the parameters of the DNN model will be changed accordingly.
In the work \cite{chenw18}, the test accuracy for unauthorized users is lower than works \cite{FanNC19, 0001ms20} and our proposed method.
The reason is that, the DNN model in work \cite{chenw18} is trained using adversarial examples. Therefore, the model only maintains high accuracy for the input adversarial examples from authorized users, but produces low accuracy for the original clean images from unauthorized users. However, in practice, users usually expect high accuracy on clean images.
Overall, the proposed \textit{ActiveGuard} method performs well in distinguishing authorized users from unauthorized users, and is implemented with low overhead.

\textbf{Users' fingerprints management.} Compared to these existing active DNN IP protection methods \cite{chenw18, FanNC19, 0001ms20}, this paper is the only work that achieves users' fingerprints management.
The existing works \cite{chenw18, FanNC19, 0001ms20} do not have the function of users' fingerprints management.
In other words, these works cannot assign each authorized user with a unique fingerprint for identity authorization, which make their methods \cite{chenw18, FanNC19, 0001ms20} unable to distinguish different authorized users (thus is not suitable for commercial digital right management applications).
On the contrary, the proposed \textit{ActiveGuard} allocates an adversarial example as the unique fingerprint of each authorized user, where each adversarial example will be classified by the DNN model as a target class $t$ with an fixed confidence $c$. In this way, each authorized user can submit the unique fingerprint to pass the identity authentication. As discussed in Section \ref{setup}, the proposed \textit{ActiveGuard} method can support a large number of authorized users.

\textbf{Ownership verification.} The DNN IP protection methods \cite{chenw18, 0001ms20} cannot verify the ownership of the DNN models.
In this paper, we propose a watermarking method for copyright verification, which aims to embed a $n$-digits watermark into the weights of DNN's convolutional layer.
Besides, compared to the method in work \cite{FanNC19}, in which the hidden parameters of passporting layers can be reversed by the pirates, the proposed method can embed watermark in discrete positions which is more flexible and stealthy, thus the embedded watermark is more difficult to be noticed.

\section{Conclusion}\label{sec5:conclusion}
This paper proposes an active DNN IP protection technique based on adversarial examples.
For the first time, the proposed \textit{ActiveGuard} method protects the copyright of DNN in three aspects: active authorization control, users' fingerprints management and copyright verification.
Specifically, to achieve authorization control, an additional control layer is designed to limit the usage of unauthorized users on a protected DNN model.
Besides, the proposed method exploits the adversarial example for users' identities management, where an adversarial example is regarded as a unique fingerprint of a user.
In this way, only the authorized user with his allocated adversarial example can pass the identity authentication.
Finally, a watermark is embedded into the weights of DNN for ownership verification.
Compared to existing active DNN IP protection works, the proposed method can achieve two more functions (users' fingerprints management and ownership verification), which are essential for practical commercial copyright management, while achieve the similar performance in active authorization control.
Experimental results on the MNIST and CIFAR-10 datasets demonstrate that, the proposed \textit{ActiveGuard} method can achieve authorization control and users' fingerprints management via unique adversarial example.
Meanwhile, the \textit{ActiveGuard} is able to embed discrete and stealthy watermark for copyright verification without affecting the normal performance of the DNN.
Further, the proposed IP protection method is demonstrated to be robust against the fingerprint forgery attack and two watermark removal attacks (model fine-tuning and pruning).

\ifCLASSOPTIONcaptionsoff
  \newpage
\fi

\bibliographystyle{IEEEtran}
\bibliography{reference}

\end{document}